\documentclass[epjST]{svjour}
\usepackage{graphicx}
\usepackage{hyperref,pbox}
\newcommand{\fig}[1]{Fig.\ \ref{#1}}
\def\eqn#1{eq.\ (\ref{#1})}

\begin{document}
\title{Evidence of dark energy in different cosmological observations}
\author{Arindam 
Mazumdar\inst{1}\fnmsep\thanks{\email{arindam.mazumdar@iitkgp.ac.in}} 
\and Subhendra Mohanty\inst{2} \fnmsep\thanks{\email{mohanty@prl.res.in}}
\and Priyank Parashari\inst{2,3}\fnmsep\thanks{\email{parashari@prl.res.in}} }
\institute{Department of Physics, Indian Institute of 
Technology Kharagpur, India - 721302 \and Theoretical Physics Division, Physical Research Laboratory, Navrangpura, Ahmedabad - 380009, India. \and Indian Institute of Technology, Gandhinagar, 382355, India}
\abstract{
The idea of a negative pressure dark energy component in the Universe which
causes an accelerated expansion in the late Universe has deep implications in 
models of field theory and general relativity.  In this article, we survey the 
evidence for dark energy from cosmological observations which started from the 
compilation of distance-luminosity plots of Type Ia supernovae. This turned out 
to be consistent with the dark energy inferred from the CMB observations and 
large scale surveys and gave rise to the concordance $\Lambda$CDM model of 
cosmology. In this article, we discuss the observational evidence for dark 
energy from Type Ia supernovae, CMB, galaxy surveys, observations of the 
Sunyaev-Zeldovich effect from clusters, and lensing by clusters. We also 
discuss the observational discrepancy in the values of $H_0$ and $\sigma_8$ 
between CMB and large scale structures and discuss if varying dark energy models 
are able to resolve these tensions between different observations.
} 
%
\maketitle
\section{Introduction}  
\label{intro}
The idea of a cosmological constant was introduced by Einstein to counteract the 
expansion in the Universe caused by normal matter as  'Einstein's Universe' was 
designed to be static.  Later the concept of the cosmological constant evolved 
to the idea that it was a form of vacuum energy with positive energy density and 
negative pressure whose effect at cosmological scales would be to cause an 
accelerated expansion of the Universe. The idea of vacuum energy has been known 
since the prediction and observation of the `Casimir effect' which arises from 
the virtual electron-positron pairs produced from the vacuum. From the point of 
quantum field theory, the existence of the cosmological vacuum energy would be 
natural, however, the problem is that the scale of vacuum energy expected in 
particle physics would be many orders of magnitude different from the energy 
density of the Universe, and this problem is usually called the `cosmological 
constant problem' \cite{Weinberg:1988cp}. Another problem from the purely 
cosmological perspective is to explain why the cosmological constant which does 
not change with the expansion of the Universe happens to be of the same order of 
magnitude as the matter in the present Universe since the two vary differently 
during the expansion history of the Universe. This `why now' problem is 
explained by coupling dark matter to normal matter so that they track each other 
over the cosmological history of the Universe.

 From the observational perspective, the first confirmation of an accelerated 
expanding Universe came from the measurement of luminosity distance plots of 
Type Ia supernovae by Riess et. al~\cite{Riess:1998cb} and Perlmutter et al.~\cite{Perlmutter:1998np}, which prompted the 
rebirth of the idea of a dark energy dominated Universe in the present epoch. 
Subsequently, the observations of CMB anisotropy and large scale structures 
confirmed the evidence of an accelerating phase for the low redshift Universe.  
The observations of supernovae luminosity, CMB and large scale structures give a 
concordance model called $\Lambda$CDM cosmology where the present Universe 
comprises about 68\% dark energy, 27\% dark matter and 5\% baryonic matter. 
However not all is well with the $\Lambda$CDM model, for example, there is a  
4-$\sigma$ discrepancy between the observation of the Hubble expansion rate 
at present epoch ($H_0$) derived from the CMB and those from the local measurements. These 
discrepancies may be addressed by the evolving dark energy models.

 In this article, we will very briefly review the different ways of measuring
 dark energy parameters and the current status of individual measurements as 
well as joint analyses. 

The Hubble parameter which determines the rate of the expansion in an FLRW Universe depends on different components of the Universe in the following way
\begin{eqnarray}\label{eq:hz}
 H(z)=H_0\sqrt{\Omega_m(1+z)^3+\Omega_{\rm DE}f(z)+\Omega_k(1+z)^2+\Omega_r(1+z)^4}\, ,
\end{eqnarray}
where $z$ represents the redshift, $\Omega_m$ is the matter density fraction, $\Omega_r$ is the radiation density fraction, $\Omega_k$ is the energy density fraction corresponds to spatial curvature of the FLRW metric and $\Omega_{\rm DE}$ is the energy density fraction of dark energy at present with
\begin{eqnarray}
 f(z)=\exp\left[3\int_0^z {1+w(z')\over 1+z'}dz'\right]\,.
\end{eqnarray}
Here $w(z)$ defines the equation of state the dark energy. We will consider three different types of dark energy scenarios;  
namely, the cosmological constant when $\Omega_{\rm DE}= \Omega_\Lambda$ and $w(z)=-1$; dark energy with a fixed non-zero equation of state parameter when $w(z) = w$ (known as $w$CDM model); and dark energy with varying equation of state parameter.    
For the last case, we consider the equation of state of dark energy to be characterized by the CPL parameterization~\cite{Chevallier:2000qy,Linder:2002et} 
\begin{eqnarray}
 w(a) = w_0 + w_a(1-a),
\end{eqnarray}
where $a$ is the scale factor of the FLRW metric of the Universe.

For all the above-mentioned cases we will study the recent observational status from different ways of measuring dark energy.
So far there are three different ways to measure the dark energy. The first one is 
through the measuring the expansion rate of the Universe directly from the 
standard candles (type Ia supernovae). After the discovery of gravitational 
waves in LIGO, the chirping frequency of the gravity wave from the binary 
mergers, which acts as a standard siren, has opened up a new window of measuring 
the distance of the binary system. If the binary system emits any electro-magnetic signal at the time of the merger, the redshift of the signal can 
provide information about the velocity of the combined system. This can 
help to infer the expansion rate of the Universe directly. However, this method 
would require the next few decades of observation to constrain dark energy 
parameters efficiently.   

The second type of the measurement is using the Baryon Acoustic 
Oscillation (BAO) scale as the standard ruler. This type of measurement 
provides the integrated history of $H(z)$ evolution. The third type of 
observations measures the growth rate of the dark matter density perturbations at 
late time. These types of observations are the large scale structure surveys 
like lensing surveys, Sunyaev-Zeldovich (SZ) surveys, galaxy surveys and, in 
future, the 21-cm surveys.                 

We organise the article in the following way. In section~\ref{sec:sne},\ref{sec:bao} and \ref{sec:growth} we describe the above mentioned three different methods of determining dark energy parameters. In these sections we presented the recent results of the observations in tabulated format (table~\ref{tab:snia} and \ref{tab:bao}). Then in two small sections, section~\ref{sec:recon} and section~\ref{sec:H0} we discuss the current issues in reconstructing dark energy equation of state from the observations and various attempts to resolve the $H_0$ tension by modifying the dark energy sector. Then we summarise the current status of all the observations in section~\ref{sec:sum}. 

\section{Measurement from the observations of type Ia supernovae}\label{sec:sne}
In a binary system, the mass of a white dwarf increases either due to accretion 
from the other star or due to merger. A type Ia supernovae (SNIa) occur when a 
white dwarf in a binary system explodes as its mass reaches the Chandrasekhar 
limit. These supernovae are the brightest of all the supernovae and they follow 
quite a similar light curve with a consistent peak luminosity. That is why we 
can use these supernovae as the standard candles to measure the distances. 
However, there are a few difficulties and problems with SNIa observations. For 
example, after the supernovae explosion, SNIa reaches the peak luminosity in a 
few weeks after that these SNIa fade away within a few months. Also, it is very 
difficult to predict a supernovae explosion event and these events take place a 
few times per millennium in a galaxy. Therefore, this makes it very difficult to 
track all the events. In addition, although most SNIa has quite a similar light 
curve, a few SNIa is either a little bit fainter or 
brighter~\cite{doi:10.1142/S0217751X00005383}. Moreover, the SNIa are observed 
in a particular band filter depending on their peak luminosity. However, some 
part of the spectrum, other than the observing filter, comes from the filter 
during observations. Therefore, we need to correct this difference in the 
spectrum to get the accurate results~\cite{doi:10.1142/S0217751X00005383}. 
Finally, the distance modulus from the observation ($\mu_{\rm obs}$) of SNIa can be obtained 
as~\cite{Scolnic:2017caz,Abbott:2018wog} 
\begin{equation}
\mu_{\rm obs} = m_B + \alpha x_1 -\beta \mathcal{C} + M_0 + \gamma G_{\rm host} + \Delta\mu_{\rm bias}\, ,
\end{equation}
where $M_0$ is the absolute magnitude of SNIa and $m_B = -2.5 \log(x_0)$ with $x_0$ being the amplitude of the light curve.
Here, $x_1$ and $\mathcal{C}$ represent the light curve width and color for each SNIa, respectively. Moreover, $\alpha$ characterize the relation between SNIa luminosity and width of SNIa light curve, and $\beta$ describes the correlation between color and SNIa luminosity. These parameters are obtained by fitting the light curve for each SNIa. Furthermore, $\gamma$ accounts for the correction due to host-galaxy stellar mass and $G_{\rm host} = +1/2$ if host-galaxy stellar mass is larger than $10^{10}$ solar mass, and $G_{\rm host} = -1/2$ if host-galaxy stellar mass is smaller than $10^{10}$ solar mass. At last, $\Delta\mu_{\rm bias}$ is determined from simulation and accounts for the selection bias.

Cosmology with supernovae depends on the luminosity distance measurement as a 
function of redshift for a number of SNIa and comparing the observed results 
with the theoretical prediction of distances in different cosmological models. 
Given a cosmological model, luminosity distance $d_{\rm L}(z)$ to a source at 
redshift $z$ can be calculated by the following relation
\begin{equation}
\label{equ:dl}
d_{\rm L}(z) = \frac{c}{H_0}(1+z)\times \left\{ \begin{array}{l l}\frac{1}{\sqrt{\Omega_k}}  \sinh \bigl( \sqrt{\Omega_k} \int_0^z \frac{dz'}{E(z')}\bigr)  & \ \quad \Omega_k > 0 \\  
\int_0^z \frac{dz'}{E(z')} & \ \quad \Omega_k = 0 \\  
\frac{1}{\sqrt{|\Omega_k|}} \sin \bigl( \sqrt{|\Omega_k|} \int_0^z \frac{dz'}{E(z')}\bigr)   & \ \quad \Omega_k < 0
\\
\end{array} \right.
\end{equation}
where $E(z)=\frac{H(z)}{H_0}$ and $H(z)$ is given by \eqn{eq:hz} and $c$ is the speed of light.
 Once we know the luminosity distance, we can calculate the distance modulus ($\mu_{th}$) for a given theoretical model as
\begin{equation}
\mu_{th} = 5 \log_{10} (d_{\rm L}(z)/10pc).
\end{equation}
Now, after comparing the theoretical and observational predictions for the distance  modulus, we can put constraints on the cosmological parameters.
\begin{table}[t]
    \centering
    \resizebox{1.0\linewidth}{!}{
    \begin{tabular}{|c|c|c|c|c|c|}\hline
    Experiment  & Model &$\Omega_\Lambda$ & $w$ & $w_0$ & $w_a$ \\ \hline
      Pantheon-SN-stat & $\Lambda$CDM &$0.716 \pm 0.012$ & - & -& - \\ 
	  Pantheon-SN-stat & $w$CDM & - & $-1.251 \pm 0.144$ & -& - \\      \hline
	  Pantheon-SN & $\Lambda$CDM &$0.702 \pm 0.022$ & - & -& - \\ 
	  Pantheon-SN & $w$CDM & - & $-1.090 \pm 0.220$ & -& - \\      \hline
	  Pantheon-SN+CMB &  $w$CDM & - & $-1.026 \pm 0.041$ & - & - \\
	  Pantheon-SN+CMB &  $w_0w_a$CDM & - & - & $-1.009 \pm 0.159$ & $-0.129\pm0.755$  \\ \hline
	  Pantheon-SN+CMB+BAO &  $w$CDM & - & $-1.014 \pm 0.040$ & - & - \\
	  Pantheon-SN+CMB+BAO &  $w_0w_a$CDM & - & - & $-993 \pm 0.087$ & $-0.126\pm0.384$  \\ \hline
	  DES-SN3YR & $\Lambda$CDM & $0.669\pm0.038$ & - &- &- \\ \hline
	   DES-SN3YR+CMB & $\Lambda$CDM & $0.670\pm0.032$ & - &- &- \\
	   DES-SN3YR+CMB & $w$CDM & - & $-0.978\pm0.059$ &- &- \\ \hline
	    DES-SN3YR+CMB+BAO & $\Lambda$CDM & $0.690\pm0.008$ & - &- &- \\
	   DES-SN3YR+CMB+BAO & $w$CDM & - & $-0.977\pm0.047$ &- &- \\
	   DES-SN3YR+CMB+BAO & $w_0w_a$CDM & - & - &$-0.885\pm0.114$ &$-0.387\pm0.430$ \\	 \hline  
    \end{tabular}
    }
    \caption{The values of dark energy related quantities from the SNIa data by pantheon~\cite{Scolnic:2017caz} and DES-SN3YR~\cite{Abbott:2018wog} samples and from the combinations of SNIa, CMB~\cite{Ade:2015xua} and BAO~\cite{Beutler:2011hx,Anderson:2013zyy,Ross:2014qpa,Alam:2016hwk} data are shown in this table. }
    \label{tab:snia}
\end{table}

In 1998, it was first discovered by Riess et. al~\cite{Riess:1998cb} using the SNIa data from Hubble space telescope (HST) observation that the Universe is expanding at an accelerating rate. In 1999, using the data of 42 SNIa, Perlmutter et al.~\cite{Perlmutter:1998np} had also found that the expansion rate of the Universe is accelerating. Since then the use of SNIa as standard candles has been of critical importance and has attracted great attention in cosmology. Over the last two decades, there have been a number of supernovae surveys by different groups which probed a large redshift range. There have been many surveys which search for SNIa in the low redshift range ($0.01 < z < 0.1$) e.g  CfA1-CfA4~\cite{Riess:1998dv,Jha:2005jg,Hicken:2009dk,Hicken:2009df,Hicken:2012zr}, the Carnegie Supernova Project (CSP)~\cite{Contreras:2009nt,Folatelli:2009nm,Stritzinger:2011qd} and the Lick Observatory Supernova Search (LOSS)~\cite{Ganeshalingam:2013mia} etc. Moreover, some surveys like the ESSENCE supernova survey~\cite{Miknaitis:2007jd,WoodVasey:2007jb,Narayan:2016fjm}, SuperNova Legacy Survey (SNLS)~\cite{Conley:2011ku,Sullivan:2011kv}, Sloan Digital Sky Survey (SDSS)~\cite{Frieman:2007mr,Kessler:2009ys,Sako:2014qmj} and Pan-STARRS survey (PS1)~\cite{Rest:2013mwz,Scolnic:2013efb} have assembled SNIa data in the redshift range $z > 0.1$. There are also some other surveys like SCP~\cite{Suzuki:2011hu}, GOODS~\cite{Riess:2004nr,Riess:2006fw} and CANDELS/CLASH~\cite{Graur:2013msa,Rodney:2014twa,Riess:2017lxs} survey which look for SNIa in high-z range ($z>1.0$). Data from these surveys had been used to constrain the cosmological parameters. Recently, Scolnic et at.~\cite{Scolnic:2017caz} have assembled the data of1042 SNIa in the redshift range from $z\sim 0.01$ to $z\sim2.0$ from PS1, CfA1-A4, CSP, SDSS, SNLS, SCP, GOODS  and  CANDELS/CLASH surveys and called it pantheon sample. They did analysis for different cosmological models with just pantheon-SN data and with the combination of pantheon-SN data and data from other cosmological probes such as CMB and BAO. The results for different cosmological models are shown in table~\ref{tab:snia}.
In addition to these, recently Dark Energy Survey Supernova Program (DES-SN) have also reported 207 SNIa in the redshift range $0.015<z<0.7$~\cite{Abbott:2018wog}. They did the analysis with a total of 329 SNIa in which they have included 122 low redshift SNIa from the literature and called this sample DES-SN3YR. The results for different cosmological models with DES-SN3YR data and data from other probes are also shown in table~\ref{tab:snia}.

Recently the authors of ref~\cite{Colin:2018ghy} have argued that the cosmic acceleration inferred from the Type Ia supernovae (more particularly the JLA data) has a scale-dependent dipolar modulation. This effect can be attributed to the bulk flow in the local Universe in which the observer is located. Therefore, the direct ``evidence'' of dark energy can be an artifact of the inhomogeneous nature of the Universe at present time. 

\section{Measurement from the imprints of baryon acoustic oscillation}\label{sec:bao}
The primordial perturbations reenter the horizon at the time of radiation dominated and the matter-dominated era. The photon-baryon fluid in which baryon and photon are strongly coupled by the Thompson scattering undergoes the acoustic oscillations due to the primordial density perturbations encountered during the expansion of the Universe. These acoustic waves set the pattern in the CMB and the galaxy distributions of the Universe. The characteristic length scale of these oscillation patterns, known as the baryon acoustic oscillation (BAO) scale, works as the standard ruler in the Universe in measuring the Hubble parameter and its evolution. It is because the length scale of BAO at the redshift of recombination can be 
expressed as~\cite{Eisenstein:2004an,Bassett:2009mm} 
\begin{eqnarray}\label{eqn:bao_sacle}
 r_s = \int_{z_{\rm rec}}^\infty {c_s\over H(z)} d z= {1\over\sqrt{\Omega_m 
H_0^2}}{2 c\over \sqrt{3 z_{\rm eq}R_{\rm eq}}}\ln 
\left[\sqrt{1+R_{\rm rec}}+\sqrt{R_{\rm eq}+R_{\rm rec}}\over 1+\sqrt{R_{\rm rec}}\right]\, ,
\end{eqnarray}
where $R_{\rm rec}$ and $R_{\rm eq}$ are baryon-photon ratio ($3\rho_b/4\rho_\gamma$) at 
the time of recombination and radiation matter equality, respectively. Here, $c_s$ is the sound speed in baryon-photon fluid, $z_{\rm rec}$ and $z_{\rm eq}$ are the redshift corresponding to the recombination and radiation matter equality epoch.  This $r_s$ can 
be estimated accurately if the redshift at recombination and the baryon density 
of the Universe is known properly. Therefore this scale serves the purpose of 
the standard ruler in estimating the size of the BAO patterns in CMB. Similarly, for the  galaxy surveys, the BAO scale is measured at $z_{\rm drag}$ which is the value of redshift when baryons and photons decouple dynamically.     

\subsection{CMB}
\begin{figure}
    \centering
    \includegraphics[width=4in]{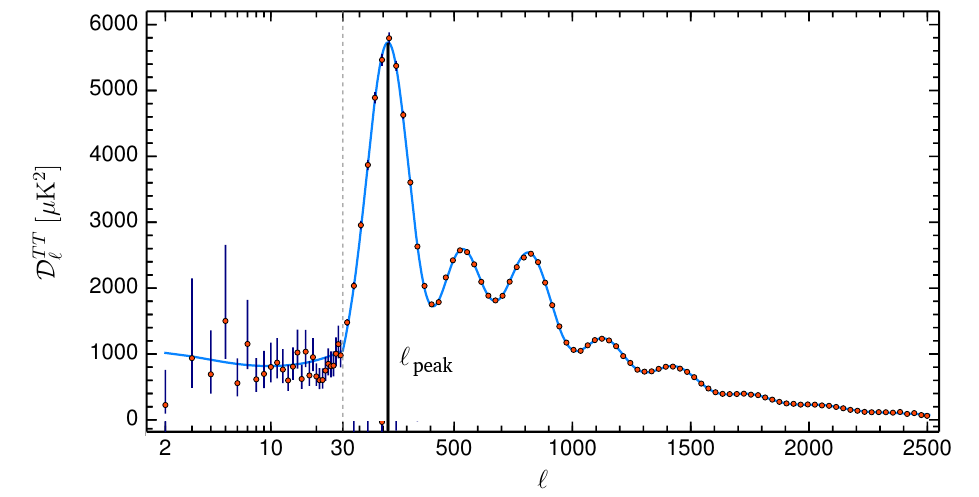}
    \caption{The figure is taken from Planck-2018 paper~\cite{Aghanim:2018eyx}. Here ${\cal D}_\ell = \ell(\ell+1)C_\ell/(2\pi)$.}
    \label{fig:cmb}
\end{figure}
Planck measurement of CMB provides the most accurate measurement of the baryon acoustic oscillation peaks. 
The CMB power-spectrum is calculated using the first order cosmological 
perturbation theory where the interaction between the baryon and photon are 
accounted for (see the equations in ref~\cite{Ma:1995ey}). The $\Delta_\ell$, 
which are the Legendre coefficients of the temperature fluctuations in CMB is 
calculated by solving these Boltzmann equations. The solution of $\Delta_\ell$ 
has an oscillatory part as well as a damping part. Further, the $\Delta_\ell$'s 
are decomposed in terms of spherical harmonics and the two-point correlation 
between the coefficients of spherical harmonics, which are written as $C_\ell$, 
are calculated. The oscillatory part in $C_\ell$  corresponds to the BAO and 
the damping is known as the Silk damping which arises from the viscosity in 
photon-baryon fluid (see \fig{fig:cmb}). The oscillatory part of the CMB can be 
approximated as $\cos(k r_s+\phi)$  where $r_s$ is defined in 
\eqn{eqn:bao_sacle}, $\phi$ is the phase factor which can depend on the effects 
of the other components of the Universe (dark matter, neutrinos) on CMB. The 
peak multipoles shown in \fig{fig:cmb} is related to the $k$ as
\begin{eqnarray}
\ell_{\rm peak} = (m\pi-\phi){D_A\over r_s} = {(m\pi-\phi)\over \theta_*} 
\end{eqnarray}
where, $D_A$ is the angular diameter distance which is given by 
\begin{eqnarray}\label{eqn:dA}
 D_A = {1\over (1+z_{\rm rec})}\int_0^{z_{\rm rec}} {c dz\over H(z)}\, .
\end{eqnarray}

The quantity $\theta_*= r_s/D_A$ is known as the angular size of BAO. Planck can measure seven BAO peaks in their CMB pattern and determines the $\theta_*$ with an 0.1\% accuracy. The information about dark energy, or background cosmology at late time in general, is inferred from CMB through this angular diameter distance. 
Whereas the Planck CMB data is extremely accurate in predicting the values of $\Omega_\Lambda$ and $w$ of $w$CDM model, the parameters of varying dark energy show high degeneracy with Planck data alone. To break the degeneracy, complementary observations like BAO and redshift space distortions are also taken into account. The results corresponding to these analyses are shown in table~\ref{tab:bao}.    

Apart from the space based surveys like WMAP or Planck, there are some ground based surveys like SPT and ACT survey which also measures the CMB temperature fluctuations in a particular portion of the sky. These experiments can also observe the BAO scales in the CMB and provide the value of the Hubble parameter. For example, the latest value of $H_0$ inferred from the ACT data alone is $67.9\pm 1.5$ $\rm km/sec/Mpc^{-1}$ and $\Omega_\Lambda$ is equal to $0.696 \pm 0.022$~\cite{Aiola:2020azj}, which are in good agreement with the Planck measurement. It has been reported that SPT data alone provides the value of $H_0$ to be $73.5 \pm 5.2$ $\rm km/sec/Mpc^{-1}$ and $\Omega_\Lambda$ is equal to $0.726 \pm 0.028$ with varying number of relativistic degrees of freedom($N_{\rm eff}$)~\cite{Balkenhol:2021eds}.

\subsection{Galaxy surveys} 
\begin{figure}\label{fig:galaxy}
\centering
 \includegraphics[width=2.5in]{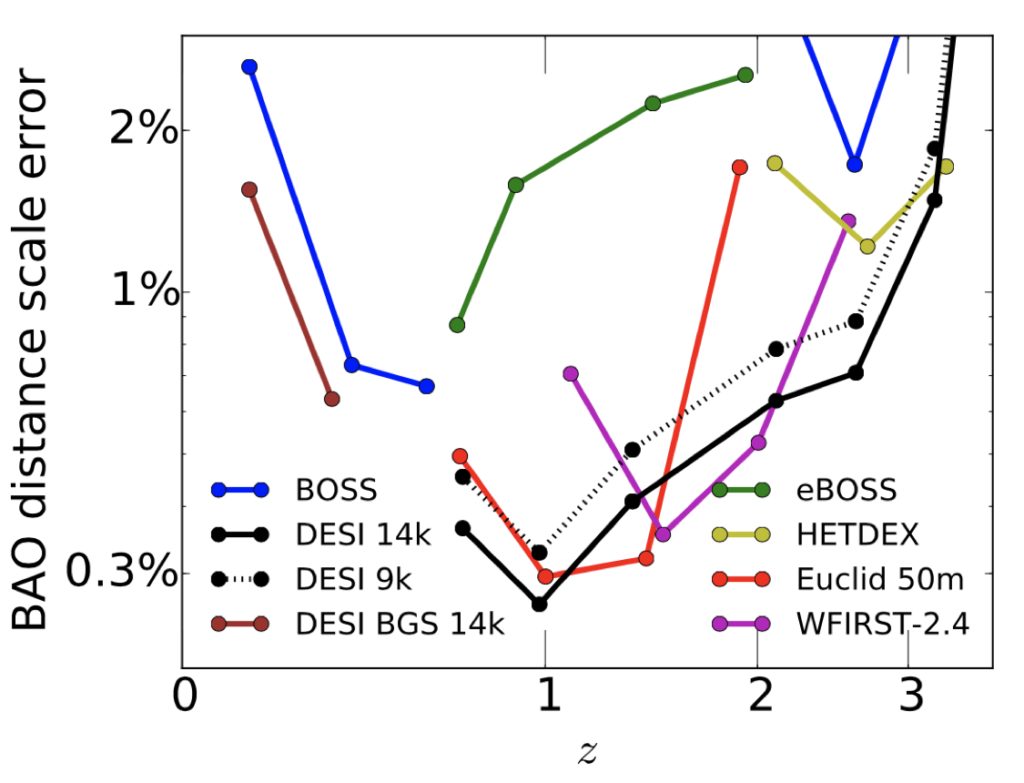}
  \includegraphics[width=2.5in]{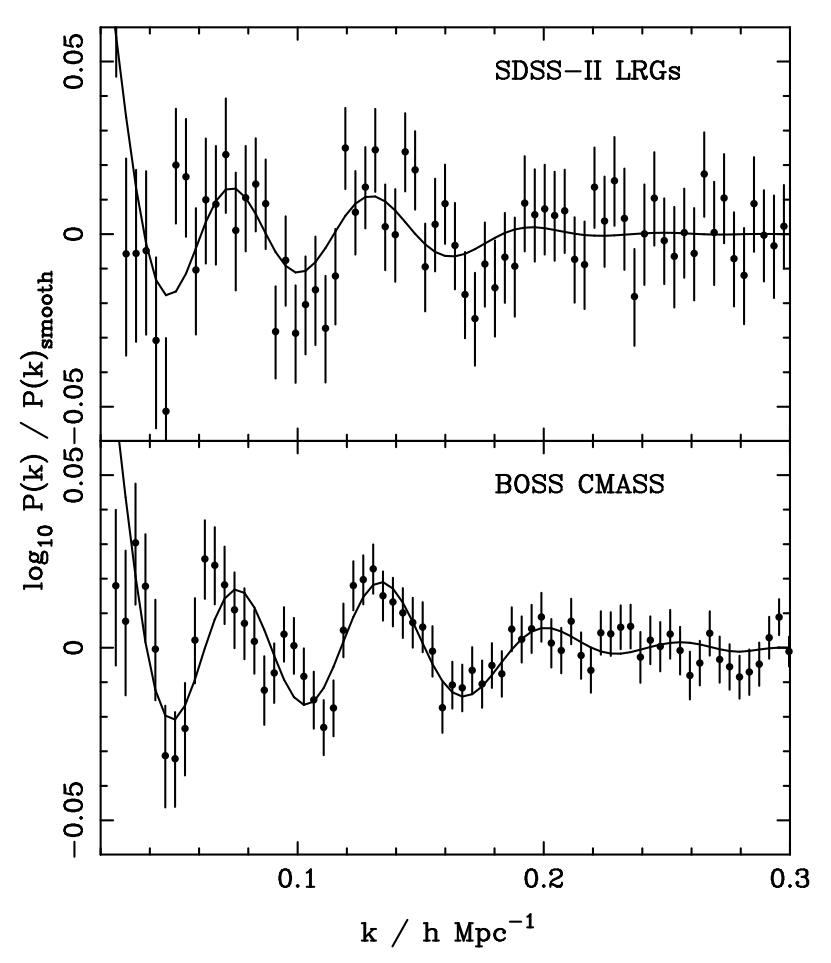}\\
  (a)\hspace*{3in}(b)
 \caption{(a) Accuracy of measurement of BAO scale increased over time with different observation mission. (figure is taken from the site of  \href{https://www.desi.lbl.gov/the-desi-science-mission/}{DESI}) (b) Signature of BAO on matter power spectrum. (figure is taken from ref~\cite{SDSS-DR9})}
\end{figure}

The galaxy surveys estimate the values of the position of the galaxies in redshift space from which the galaxy distribution in real space is calculated. The power spectrum of the galaxy field is  
\begin{eqnarray}
 P_g(k)= b_1^2 P(k)\, ,
\end{eqnarray}
where $b_1$ is the linear bias factor and $P(k)$ is the cold dark matter power spectrum. In \fig{fig:galaxy}-(b) the matter power spectrum of the observed galaxy field is plotted after subtracting the smooth theoretical power spectrum from it. The left-over power spectrum shows the oscillation pattern of BAO.

From these BAO patterns, the galaxy surveys measure
\begin{eqnarray}
d_z = {r_s(z_{\rm drag})\over D_V(z)}\, ,  
\end{eqnarray}
where, 
\begin{eqnarray}
 D_V(z) = \left[(1+z)^2 D_A(z)^2 {c\, z\over H(z)}\right]^{1/3}\, .
\end{eqnarray}
Here, $D_A(z)$, the angular-diameter-distance whose expression is given in \eqn{eqn:dA} in which $z_{\rm rec}$ has to be substituted by the redshift of the galaxies.

The experiments that measure BAO in galaxy power-spectrum are  2dF Galaxy survey~\cite{Cole:2005sx}, SDSS and 
BOSS~\cite{SDSS-DR9,Percival:2009xn,SDSS-DR9,Alam:2016hwk}, Wiggle-Z~\cite{Blake:2011en}, 6dF galaxy survey~\cite{Beutler:2011hx}.
The upcoming galaxy surveys are Euclid~\cite{euclid2011} and DESI~\cite{desi:2013}. These galaxy surveys not only provide the BAO pattern but also provides the information of growth factor from redshift space distortion which we will be discussing later in this article.

\begin{table}[]
    \centering
    \resizebox{1.0\linewidth}{!}{
    \begin{tabular}{|c|c|c|c|c|}\hline
    Experiment  & $\Omega_\Lambda$ & $w$ & $w_0$ & $w_a$ \\ \hline
      Planck (TT,TE,EE, lowE) & $0.6847 \pm 0.0073$ & - & -& - \\ \hline
      Plnack+SNE+BAO & - & $-1.028 \pm 0.031$ & $ -0.957 \pm 0.080$ & $-0.29^{+0.32}_{-0.26}$\\ \hline
      Plnack+BAO/RSD+WL & - & - & $-0.76 \pm 0.20$ & $-0.72^{+0.62}_{-0.54}$\\ \hline
      \pbox{1in}{Planck+JLA+WiggleZ\\+CFHTLens+SDSS-DR12~\cite{Zhao:2016das}} & 
- & - & $-0.96 \pm 0.10$ & $- 0.12 \pm 0.32$ 
\\ \hline
      Planck+SDSS-DR12~\cite{Zhao:2016das} & - & - & 
$-1.2 \pm 0.32$ & $- 0.33 \pm 0.75$
\\ \hline
     Planck-SZ + BAO & - &$-1.01 \pm 0.18$& - & - \\ \hline
      \pbox{1in}{CFHTLens+WMAP7\\+BOSS+HST~\cite{Heymans:2013fya}\\ (for flat Universe)}& $0.729\pm 0.010$& $-1.02 \pm 0.09$ & - & - \\ \hline
      DES Y1 &$0.733^{+0.030}_{-0.017}$ & $0.82$ & - & - \\ \hline
      DES+Planck+BAO+ SNe & $0.702\pm 0.007$ & $-1^{+0.05}_{-0.04}$ & - &- \\ \hline
    \end{tabular} }
    \caption{The values of dark energy related quantities from the observations 
of baryon acoustic oscillation, RSD, SZ and lensing. In the second and third 
column ``Planck" means ``Planck TT,TE,EE+lowE+lensing".}
    \label{tab:bao}
\end{table}

\section{Measurement from the growth rate of large scale structures}\label{sec:growth}
The growth rate of density perturbations of cold dark matter in the late time 
of the evolution of the Universe provides a robust probe to the existence of 
dark energy and its equation of state. The linear growth factor of the dark 
matter density perturbations can be written 
as~\cite{Heath-77,Eisenstein:1997ij,Hamilton:2000tk} 
\begin{eqnarray}
 G(z) =  5{\Omega_m E(z)
 \over 2}\int_Z^\infty {(1+z)dz\over E(z')^3}
\end{eqnarray}
where $E(z)= H(z)/H_0$ and $E(z)$ contains the by dark matter density ($\Omega_m$) and dark energy 
density ($\Omega_\Lambda$). The 5/2 factor in the above equation is a 
normalization factor to make $G$ equals to one at $a=1$. In the case of the matter dominated Universe,  
the $G(z)$ becomes equal to ($1/1+z$) or $a$. However, in the case of dark energy dominated 
Universe the situation changes and the growth factor is often denoted by a quantity
\begin{eqnarray}
 f = {d\ln G \over d \ln a}\, .
\end{eqnarray}
Therefore, the deviation of $f$ from the value of one provides the information 
about dark energy. It is customary to fit the $G(z)$ numerically and express it 
in terms of a monomial function of dark matter density fraction $\Omega_m(z)$  as 
\begin{eqnarray}
G(z)=\Omega_m(z)^\alpha\, .    
\end{eqnarray}
The reason behind it is it makes the calculation of higher order perturbation analytically possible.
\subsection{Redshift space distortion}
Redshift space distortion imprints unique patterns on the distribution of the 
tracer field (galaxy or atomic hydrogen) which changes the position  of the 
tracer field in redshift space depending on their peculiar velocities. Within 
the fully virialized 
objects, like the large halos, the peculiar velocities of the galaxies are much 
higher and random. However, on the larger scales the peculiar velocities are 
correlated to the density contrast and their magnitude can be predicted using 
the growth function in the ambit of cosmological perturbation theory. Therefore, 
the redshift space distortion provides an independent measurement of the growth 
factor along with the amplitude of the power spectrum. 

In the redshift space, the power-spectrum loses its isotropic nature and 
its value depends on the direction of $\vec{k}$-vector with respect to the line 
of sight. In first order perturbation theory under plane parallel assumption the redshift space power spectrum can be written as 
\begin{eqnarray}
 P^s_g(k,\mu)= b_1^2 (1+\beta\mu^2)^2 P(k)\, ,
\end{eqnarray}
where $\mu$ is the cosine of the angle between the $\vec{k}$-vector and the line 
of sight. To quantify this distortion it is customary to decompose the 
anisotropy in terms of spherical harmonics. In that case the monopole of the 
redshift space power-spectrum  is amplified by the Kaiser factor which can be
written as~\cite{Kaiser:1987qv} 
\begin{eqnarray}
 P_g^0(k)=b_1^2\left(1+{2\over3}\beta+{1\over 5}\beta^2\right)P(k)\, .
\end{eqnarray}
Here, $\beta=f/b_1$. Similarly, the quadrupole and hexadecapoles can also be calculated.
However, the model power-spectrum considered by BOSS-DR12 is even more 
complicated which 
contains not only the extension of the Kaiser model but also the higher order terms~\cite{Beutler:2016arn}, 
\begin{eqnarray}
 P_g^s(k,\mu)&=& \exp[-(f k \mu \sigma_v)^2]\lbrace P_{g,\delta\delta}(k)+ 2 f\mu^2 P_{g,\delta\theta}(k) + 
 f^2\mu^4 P_{g,\delta\theta}(k)+\nonumber \\ &&{\rm other\,\, correction\,\, terms}\rbrace\, .
\end{eqnarray}
The exponential term defines the ``Finger of God" effect due to the random velocities of the galaxies on the small scales. Here, $\theta$ is the divergence of the velocity fields of the galaxies, $\sigma_v$ is the standard deviation of the velocity distribution of galaxies. 
Measurement of the multipoles of this power spectrum provides the value of $f(z) \sigma_8(z)$. For example the latest BOSS result provides
the value of $f\sigma_8 = 0.289^{+0.085}_{-0.096}$ for $z_{\rm eff} = 
0.85$~\cite{deMattia:2020fkb}. Therefore, if $\sigma_8$ is measured from other 
complementary observation, then RSD can constrain the growth quite effectively. 
So, when Planck CMB data alone cannot constrain dark energy parameters 
effectively, RSD is incorporated in the analysis which provide the value of 
$w_a$ and $w_0$ with reasonable accuracy (see table~\ref{tab:bao}). In an 
earlier analysis of {\bf BOSS-DR 11} power spectrum along with Planck provided 
the value of CPL parameters~\cite{Upadhye:2017hdl} to be $ w_0 = 
-0.87^{+15}_{-16}$ and $\, w_a = -0.61^{+0.76}_{0.61}$.

However, the degeneracy between $\sigma_8$ and $f$ cannot be broken with the 
observation of the redshift space power spectrum alone. For that purpose, it is 
essential to measure the bi-spectrum in the redshift space. The multipole 
moments of redshift space bi-spectrum for the second order perturbation theory 
has been studied in ref~\cite{Scoccimarro:1999ed,Mazumdar:2020bkm}. However, 
there do not exist any constraints on dark energy parameters using redshift 
space bispectrum as of now.

\subsection{Lensing surveys}
\begin{figure}\centering
 \includegraphics[width=3in]{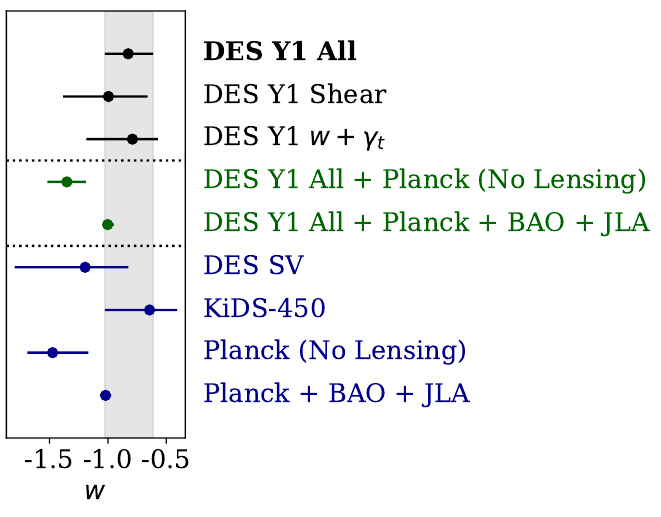}
 \caption{Best-fit values of $w$ for $wCDM$ model from the lensing surveys (figure taken from 
ref~\cite{Abbott:2017wau}).}\label{fig:des}
\end{figure}

These surveys observe the lensing features, mainly the B-modes, produced by the large scale structures on CMB and traces back the lensing potential($\phi$) from that. The power spectrum of lensing potential  ($C_\ell^{\phi\phi}$) can be theoretically obtained from the power-spectrum of the gravitational potentials ($P_\Psi$)~\cite{Lewis:2006fu,Smith:2008an}.
\begin{eqnarray}
 C_\ell^{\phi\phi} = {8 \pi^2\over \ell^3}\int_0^{z_{\rm rec}} {dz\over H(z)}\chi(z)\left(\chi(z_{\rm rec}) - \chi(z)\over \chi(z_{\rm rec})  \chi(z)\right)^2 P_\Psi(z,k= \ell/\chi(z))
\end{eqnarray}
Here $\chi(z)$ is the comoving distance at $z$. The power spectrum for gravitational potential is related to the matter power spectrum through the Poisson's equation as
\begin{eqnarray}
 P_\Psi(z,k) = {9 \Omega_m(z)^2 H(z)^4 \over 8\pi^2}{P(z,k)\over k} 
\end{eqnarray}
The non-linear matter power-spectrum is calculated using the {\tt 
HaloFit}~\cite{Smith:2002dz} extrapolation and linear matter power-spectrum as 
the input. The dependence of the linear matter spectrum on the growth function 
is discussed in the last subsection. Therefore, the dependence of the lensing 
potential with the growth factor is evident. The most important lensing surveys are 
CFHTLens~\cite{Heymans:2013fya,Erben:2012zw}, KiDs~\cite{Hildebrandt:2016iqg}, 
Planck Lensing survey~\cite{Aghanim:2018oex} and DES~\cite{Abbott:2017wau}. The 
constraints on the equation of state in $w$CDM model from all the lensing 
surveys are shown in \fig{fig:des}. The values of dark energy parameters 
constrained using lensing surveys are listed in table~\ref{tab:bao}.

\subsection{Sunyaev Zeldovich surveys}%
These surveys aim at counting the number of galaxy clusters from their SZ 
effect on CMB. In SZ effect~\cite{Sunyaev:1970eu,Sunyaev:1972eq,Sunyaev:1980vz} 
the black-body spectrum of the CMB gets distorted due to the inverse Compton 
scattering of CMB photon by the free electrons present in the intergalactic 
medium of the halo containing a galaxy cluster. Among the two types of SZ 
distortion known as $\mu$ distortion and $y$ distortion, it is the $y$ 
distortion that is mainly used to count the number of the clusters in the CMB 
surveys like space based Planck and ground based SPT and ACT survey. 
The masses of the observed clusters in SZ survey are assigned from the lensing survey. 
In this way, SZ surveys provide the number of halos in a given mass range and redshift 
range in the Universe. Theoretically the number density of halos of a given mass range can be calculated from the halo mass function.
Although the simplest form of halo mass function, known as the Press-Schechter 
mass function can be calculated analytically, more accurate halo mass function 
requires $N$-body simulation. The most popular numerically fitted halo mass 
function is the Tinker~\cite{Tinker:2008ff} halo mass function which considers 
dark energy as cosmological constant. Whether different dark energy models will 
change the halo mass function or not is still an unresolved matter in cosmology. 
In general, it is assumed that halo mass function should depend on $\Omega_m$ and 
$\sigma_8$ only, which is known as the universality of halo mass function. 
However, it has been reported that different models of dark energy break the 
universality~\cite{Courtin:2010gx}. Recently it has been again claimed that 
universality can be restored by rescaling some variable~\cite{Despali:2015yla}. 
Therefore, SZ surveys provide mainly the values of $\Omega_m$ and $\sigma_8$ 
and it cannot distinguish between the different dark energy models in an 
efficient way. However, when the measurements of $\sigma_8$ and $\Omega_m$ from 
an SZ survey is combined with other experiments it can provide reasonably good 
constraints on dark energy parameters. Planck SZ and BAO joint analysis provides 
$w= -1.01 \pm 0.18$ for $w$CDM model~\cite{Ade:2015fva}. A recent joint 
analysis of DES and SPT has provided $w= -1.76^{+0.33}_{-0.46}$ for $w$CDM 
model along with varying neutrino mass~\cite{Costanzi:2020dgw}.   

\section{Reconstruction of dark-energy equation of state from observations:}\label{sec:recon}
\begin{figure}
    \centering
    \includegraphics[width=0.5\linewidth]{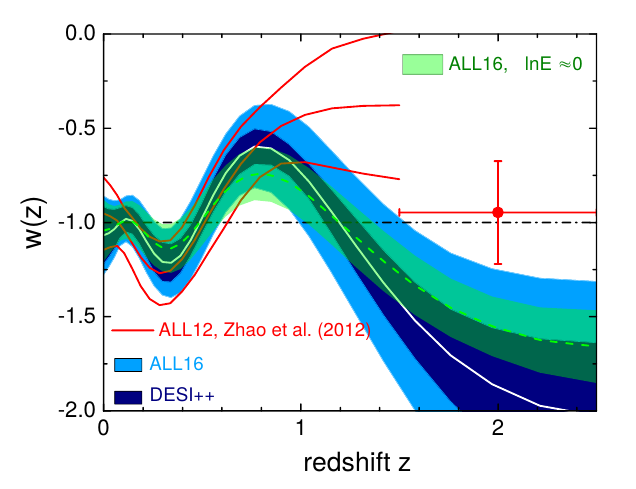}
    \caption{Reconstruction of $w(z)$ from different data sets (11 BAO data sets and CFHTLens) shows that a dark energy behaves as a phantom field for significant range of redshift. The figure is taken from ref~\cite{Zhao:2017cud}.}
    \label{fig:w_z}
\end{figure}

From the above discussions, we find that the most accurate determination of dark 
energy parameters come from the BAO and the measurement of the expansion of the 
Universe through the standard candles. Different galaxy surveys can provide BAO 
scales at different $z$ values. As well as CMB can provide the estimation of 
angular scale at $z_{\rm rec}$. Therefore, in recent years there has been series 
of attempts to reconstruct the $H(z)$ from the BAO values of different 
observations. The reconstruction of $H(z)$ also provides the reconstruction of the 
dark energy equation of state 
($w(z)$)\cite{Zhao:2012aw,Zhao:2017cud,Dai:2018zwv,Dinda:2019mev}. The 
reconstruction depends highly on the parametrization of the $w(z)$. However, 
there is a common finding among most of the studies.  For some values 
of $z$, $w(z)$ becomes less than $-1$. Any field which has an equation of state 
less than $-1$ is called the phantom field and therefore this type of dark 
energy is known as phantom dark energy. Some studies even show an oscillating 
feature in the dark energy equation of state (see \fig{fig:w_z}).
 
\section{$H_0$ tension and Dark energy:}\label{sec:H0}
The value of Hubble parameter inferred from the CMB and BAO measurements using 
the $\Lambda$CDM cosmology strongly disagrees with the Hubble value obtained 
from the direct measurement of HST~\cite{Efstathiou:2013via}. This tension has 
opened up many possibilities of modifying the dark energy sector to resolve this 
tension. An incomplete list of such models includes early dark 
energy~\cite{Karwal:2016vyq,Poulin:2018dzj,Poulin:2018cxd,Smith:2019ihp}, interacting dark 
energy~\cite{DiValentino:2017iww}, dynamical dark 
energy~\cite{DiValentino:2017zyq} etc. However, it has been also argued that no 
model can resolve the $H_0$ tension by just modifying the dark energy dynamics 
in late time~\cite{Guo:2018ans}. 
Here, we briefly review some of the features these models
\begin{itemize}
    \item{\bf Early dark energy} This solution proposes that some component of dark energy in the early times ($z \ge 3000$) behaves as a cosmological constant and makes a significant contribution to the energy density of the Universe. Then it decays down to radiation or some other component. This kind of models essentially modifies the growth rate of perturbations in the early times by modifying the background expansion rate for a certain period. However, there remain some issues with the early dark energy as it eases the tension between CMB and direct measurement of Hubble value but cannot resolve the tension between the inferred values from galaxy surveys and direct observation~\cite{Hill:2020osr,Ivanov:2020ril}. To resolve these issues further ``new early dark energy" models are also proposed~\cite{Niedermann:2020qbw,Niedermann:2020dwg}.  
    \item{\bf Interacting dark energy} In these models interaction between dark matter and dark energy are considered where the energy exchange is proportional to the four-velocity of the dark matter~\cite{Valiviita:2008iv,Vergani:2008jv,Honorez:2010rr}. These models ease the tension between Planck and HST data. However, these models fail to resolve the tension between the BAO data and HST data~\cite{DiValentino:2019ffd}.  
    \item{\bf Dynamical dark energy} As discussed earlier these types of model consider the dark energy equation of state to vary with time and CPL parameterization is the most popular way to of quantifying that variation. It has been shown that a joint analysis of Planck+HST data provides~\cite{DiValentino:2017zyq} $w_0=-1.39^{+0.39}_{-0.32}$ and $w_a=-0.2^{+0.8}_{-1.6}$ and $H_0$ to be $73.9 \pm 2.0$. However, when BAO+Planck data are considered the analysis provides a very low value of $H_0$. Therefore, it cannot be claimed as a solution to the $H_0$ tension.   
\end{itemize}

In spite of all these attempts, the very essence of $H_0$ tension is 
still intact. No other cosmological model except $\Lambda$CDM can fit the Planck 
CMB data alone with a better chi-squared value. The tension in between the 
different data-sets are still there and any solution which reduces the tension 
in $H_0$ increases the tension in $\sigma_8$ or other parameters.

\section{$\sigma_8$ tension and Dark energy:}\label{sec:S8} 
 Moreover, it has also been reported that there is a mismatch between the value 
of $\sigma_8$, the r.m.s fluctuations of density fluctuations at 8 
$h^{-1}$Mpc$^{-1}$, inferred from CMB fitted parameters under the 
$\Lambda$CDM framework and 
LSS observations~\cite{Joudaki:2016kym,Mohanty:2018ame,DiValentino:2020vvd}. This is commonly known as $\sigma_8$ tension. There have been a few attempts to resolve this 
tension by modifying the dark energy physics which includes interacting dark energy, dynamical dark energy 
etc~\cite{Pourtsidou:2016ico,Salvatelli:2014zta,Yang:2014gza,Gomez-Valent:2017idt,Gomez-Valent:2018nib,Lambiase:2018ows,DiValentino:2019ffd,DiValentino:2019jae,Davari:2019tni,Parashari:2021qjg}. In refs.\cite{DiValentino:2019ffd,DiValentino:2019jae}, interaction between the dark energy and dark dark matter have been explored and they show that $\sigma_8$ tension significantly reduces in such models. In ref.~\cite{Davari:2019tni}, minimally and non-minimally coupled scalar field, which can act as the possible alternatives for dark energy, have been proposed to ease the $\sigma_8$ tension. In ref.~\cite{Lambiase:2018ows}, dynamical dark energy (CPL parameterization~\cite{Chevallier:2000qy,Linder:2002et}) and $f(R)$ gravity model for dark energy have been analyzed and they find that $\sigma_8$ tension slightly decreases in the case of dynamical dark energy, whereas it worsens in case of $f(R)$ gravity model for dark energy. Furthermore, there are a few works that explore the running vacuum models as a resolution to the $\sigma_8$ tension~\cite{Gomez-Valent:2017idt,Gomez-Valent:2018nib}. The interested reader can see ref.~\cite{DiValentino:2020vvd} for a brief review on the current status of $\sigma_8$ tension.

\section{Summary}\label{sec:sum}
In this article, we have briefly reviewed the methods and the current status of measuring dark energy parameters from different observations. The best measurement comes from the baryon acoustic oscillation scales of Planck CMB data. For the case of the $\Lambda$CDM model experiments like the recent lensing observations (DES) or Planck data can provide significantly tight constraint by themselves. However, for the case of the $w$CDM model or dynamically varying dark energy models still, no single observation can constrain the parameters. In these cases, the joint analyses help to resolve the degeneracy between the parameters. For example, the RSD data of BOSS-DR12 can provide only the $f\sigma_8$ combination in a particular $z_{\rm eff}$. SZ observations can provide a contour in the $\Omega_m$-$\sigma_8$ plane. Lensing observations also provide the best constraints on the $\Omega_m$-$\sigma_8$ plane. BAO from the galaxy surveys or Planck shows huge degeneracy in CPL parameters when analysed alone. But When Planck data is combined with RSD and lensing data it narrows down $w_0$ to $-1$ and $w_a$ to 0.  

Different observations, although helps to break the degeneracy in the more complicated models, create tensions for the simplest models.
The BAO data at different $z$ values from different observations does not favor simple $\Lambda$CDM cosmology. Rather the recent reconstruction of the dark energy equation of state from these data sets has shown some hint of phantom dark energy for some particular ranges of redshift values. Similarly, the Hubble tension between the CMB and the direct detection of $H_0$ also opened up the scope of exploring different dark energy models. 


\bibliographystyle{JHEP}
\bibliography{reference}

\end{document}